# An Analysis of Energy Consumption on ACK+Rate Packet in Rate Based Transport Protocol

P.Ganeshkumar
Department of IT
PSNA College of Engineering & Technology,
Dindigul, TN, India, 624622
p_ganeshkumar@rediffmail.com

K.Thyagarajah
Principal
PSNA College of Engineering & Technology,
Dindigul, TN, India, 624622
drkt52@gmail.com

*Abstract*— **Rate based transport protocol determines the rate of data transmission between the sender and receiver and then sends the data according to that rate. To notify the rate to the sender, the receiver sends ACK+Rate packet based on epoch timer expiry. In this paper, through detailed arguments and simulation it is shown that the transmission of ACK+Rate packet based on epoch timer expiry consumes more energy in network with low mobility. To overcome this problem, a new technique called Dynamic Rate Feedback (DRF) is proposed. DRF sends ACK+Rate whenever there is a change in rate of ±25% than the previous rate. Based on ns2 simulation DRF is compared with a reliable transport protocol for ad hoc network (ATP) .**

## I. INTRODUCTION

In this paper, the design of new technique called Dynamic rate feedback (DRF), which minimizes the frequency of ACK+Rate packet transmission for rate based transport protocol in ad hoc network is focused. Ad hoc network is dynamically reconfigurable wireless network that does not have a fixed infrastructure. The characteristics of ad hoc network are completely different from that of the wired network. Therefore TCP protocol which is designed originally for wired network cannot be used as such for ad hoc network. Several studies have focused on the transport layer issues in ad hoc network. Research work have been carried out on both studying the impact of using TCP as the transport layer protocol and improving its performance either through lower layer mechanisms that hide the characteristics of ad hoc network from TCP, or through appropriate modifications to the mechanisms used by TCP [1-8]. Existing approaches to improve transport layer performance over ad hoc networks fall under three broad categories [9] (i) Enhancing TCP to adopt to the characteristics of ad hoc network (ii) Cross layer design (iii) A new transport protocol tailored specifically for ad hoc network environment. TCP ELFN proposed by Holland et al [10], Atra framework proposed by Anantharaman et al [11], Ad hoc transport protocol (ATP) proposed by Sunderasan et al [12] are examples protocols of the three categories respectively. Transport layer protocols tailored specifically for ad hoc network are broadly classified in to (i) Rate based transport protocol. (ii) Window based transport protocol. In rate based transport protocol the rate is determined first and then the data are transmitted according to that rate. The intermediate node calculates the rate of data transmission. This rate is appended in the data packet and it is transmitted to the receiver. Receiver collates the received rate from the intermediate node and sends it to sender the along with ACK. This ACK+Rate packet is transmitted based upon epoch timer expiry. If the epoch timer is set to 1 second, then for each and every 1 second the receiver transmits ACK+Rate packet to the sender. In this paper, the epoch timer based transmission of ACK+Rate packet and its problems are discussed. The frequency of ACK+Rate packet transmission with respect to the energy consumption, mobility and rate adaptation is presented. Frequency of rate change with respect to mobility and its result is also presented. Energy consumption with respect to simulation and its results are found out for various mobility speeds.

## II. RELATED WORK

In this paper, the focus is based on the proposals that aim to minimize the frequency of the ACK packets transmitted. Jimenz and Altman [13] investigated the impact of delaying more than 2 ACKs on TCP performance in multi hop wireless networks and they proved through simulation that encouraging result have been obtained. Johnson [14] investigated the impact of using extended delayed acknowledgement interval on TCP performance. Allman [15] conducted an extensive simulation evaluation on delayed acknowledgement (DA) strategies. Most of the approaches aims only on the ACK packets of window based transmission. In contrast to window based transmission, rate based transmission is another classification which falls under transport layer protocols as mentioned in section 1. As compared to window based transmission, rate based transport protocols aid in improving the performance in the following 2 ways [12] (i) Avoid the draw back due to burstiness. (ii) the transmission are scheduled by a timer at the sender, therefore the need for self clocking through the arrival of ACK is eliminated. The latter benefit is used by rate based protocols to decouple congestion control mechanism from the reliability mechanism. The protocols that fall under rate based transmission are as follows. Sunderasan et al [12] proposed ATP: a reliable transport protocol for ad hoc network. Kaichen et al [16] proposed an end to end rate based flow control scheme called EXACT. Ashishraniwala et al [17] designed a link layer aware reliable transport protocol called LRTP. Danscofield [18] in his thesis presented a protocol called a hop by hop transport control for multi hop wireless network. Nengchungwang et al [19] have proposed improved transport protocol which uses fuzzy logic control. Ganeshkumar et al [20,21] have studied ATP and designed a new transport protocol called PATPAN. In all the rate based transport protocol mentioned above, the rate (congestion information) is transmitted by the intermediate node to the receiver during the transmission of data packets. The receiver collates the congestion information and notifies the same to the sender along with ACK. The sender adjusts the rate of data packet transmission according to the rate (congestion information) received from the receiver. The ACK+Rate feedback packet is transmitted periodically or based on epoch timer. The granularity of setting epoch timer and the frequency of ACK+Rate feedback packet transmission highly influences the performance of the protocol. In window based transport protocol huge amount of work is done to minimize the frequency of ACK packet transmission. The literature survey on the related work of rate based transport protocol clearly depicts that until now research work have not been carried out in minimizing the frequency of ACK+Rate feedback packet transmission. Therefore this motivated us to study the behaviour of well known rate based transport protocol ATP [12] and further explore deep in to the frequency of ACK + Rate feedback packet transmission.

## III. BACKGROUND

### A. Ad Hoc Transport Protocol (ATP)

Ad hoc transport protocol (ATP) is a protocol designed for ad hoc wireless networks. It is not based on TCP. ATP differs from TCP in many ways: ATP uses coordination between different layers, ATP uses rate based transmissions and assisted congestion control and finally, congestion control and reliability are decoupled in ATP. Like many TCP variants, ATP also uses information from lower layers for many purposes like estimating of initial transmission rate, congestion detection, avoidance and control, and detection of path breaks. ATP obtains network congestion information from intermediate nodes, while the flow control and reliability information are obtained from the ATP receiver.

The ATP uses a timer-based transmission where the rate is dependent on the congestion in the network. As packets travel through the network, intermediate nodes attach the congestion information to each ATP packet. This congestion information is expressed in terms of weighted average of queuing delay and contention delay experienced by the packets at intermediate node.

The ATP receiver collates this information and sends it back to the sender in the next ACK packets, and the ATP sender can adjust its transmission rate based on this information. During the establishment of the connection, the ATP sender determines the initial transmission rate by sending probe packet to the receiver. Each intermediate node attaches network congestion information to the probe packet and the ATP receiver replies to the sender with an ACK packet containing relevant congestion information. In order to minimize control overhead, ATP uses connection request and ACK packets as probe packets.

ATP increases the transmission rate only if the new transmission rate (R) received from the network is beyond a threshold (x) greater than a current rate (S), e.g. if R>S(1+x) then the rate is increased. The transmission rate is increased only by a fraction (k) of the difference between two rates, i.e.: S=S+(RS)/k, this kind of method avoids rapid fluctuations in the transmission rate. If the ATP sender has not received ACK packets for two consecutive feedback periods, it significantly decreases the transmission rate. After a third such period, connection is assumed to be lost and the ATP sender moves to the connection initiation phase where it periodically generates probe packets. When a path break occurs, the network layer sends an explicit link failure notification (ELFN) packet to the ATP sender and the sender moves to the connection initiation phase. The major advantage of ATP is the avoidance of congestion window fluctuations and the separation of the congestion control and reliability. This leads to a higher performance in ad hoc wireless networks. The biggest disadvantage of ATP is incompatibility with a traditional TCP, nodes using ATP cannot communicate directly with the Internet. In addition, fine-grained per-flow timer used at the ATP sender may become a bottleneck in large ad hoc wireless networks.

## IV. DYNAMIC RATE FEEDBACK (DRF)

### A. Design issues

In rate based transport protocols, the intermediate node sends the congestion information (rate) to the receiver. The congestion information is usually appended with the data packet. The receiver collates the congestion information and sends it to the sender. The sender finds out the rate according to the received congestion information and starts to transmit the data. According to the concept of IEEE 802.11 MAC, at any particular instant only one node make use of the channel, even though the channel is common for all the nodes lying in the same contention area. The receiver sends transport layer ACK for the data which it have received. Since MAC uses shared channel, the data packet and ACK packet will contend to occupy the channel. This reduces the number of data packet send. To address this issue SACK is used in the transport layer. In TCP SACK, the ACK will be send for every 3 packets. In ATP SACK, the ACK will be send for every 20 packets or less than 20 packets. In ATP, to trigger the process of sending ACK, epoch timer is used. This epoch timer has a fixed value. After each epoch timer expires, ACK will be send

The congestion information which the intermediate node sends to the receiver is transmitted by the receiver to the sender along with ACK. The ACK is triggered by epoch timer expiry. Therefore for each and every epoch timer expiry the congestion information (delay/rate) is send to the sender.

Even though the rate is determined on the fly and dynamically through out the session, it is notified to the sender only once per epoch. Therefore the granularity of epoch affects the performance of the protocol. If the epoch time is very short, then ACK packet transmitted per unit time will become more. This unnecessary traffic in the reverse path creates lot of problems such as high energy consumption leading to poor network life time and throughput reduction in the forward path. If the epoch time is very large, then ACK+Rate packet transmitted per unit time will be less. Due to this the rate changes which occur within the epoch will not be notified to the sender promptly. This causes the sender to choose rate that are much lower than necessary, resulting in periodic oscillations between high and low transmission speeds.

In ad hoc network, the topology of the network changes dynamically which result in frequent rate changes. If the sender is not notified with frequent rate changes, then the sender will choose a rate which does not appropriately matches with the characteristics of network at that particular instant. Due to this the sender may choose a rate that is much lower or higher than necessary resulting in periodic oscillation between high and low transmission speeds.

The epoch timer plays an important role in terms of throughput in the forward path, energy consumption and congestion in the network. According to the characteristics of the ad hoc network choosing a constant epoch timer value for the entire period of operation of protocol does not hold good. Therefore in the proposal the ACK+Rate packet is transmitted by the receiver to the sender when ever there is 25 percent rate change than the previous rate. If the receiver finds a rate which is 25 percent more or less than the previous rate, then it transmits the ACK+Rate to the sender. This procedure is termed as Dynamic Rate Feedback(DRF). According to this technique the rate is notified to the sender when ever there exist a 25 percent rate change, eliminating the concept of ACK+Rate packet transmission based on epoch timer expiry.

### B. Independence on ACKs

In rate based transport protocols, the rate is notified to the sender along with ACK. The receiver sends ACK+Rate to the sender, the sender adopts its date transmission according to the received rate. In window based transmission, the reception of ACK triggers the data transmission. So if ACK is late or if only few ACK is received per unit time, then only less number of data packets will be transmitted. This decreases the throughput and performance of the protocol. In rate based transport protocols, the reception of ACK does not trigger the data transmission. The data is transmitted only based on the received rate. In DRF while using slow speed mobile topology, transmission of ACK+Rate packet is limited. This does not affect the number of data packet transmitted in the forward path. In low mobility network, the frequency of rate changes will be minimum. This does not cause any side effect except for which the sender should not discard the buffered data transmitted already until ACK is received for the same. DRF does not affect the data recovery through retransmission. If any data is lost, then it is the indication of congestion, route failure. This causes change in rate. If rate change occurs, then ACK+Rate will be transmitted. From the ACK, the lost data packet can be found out and the same can be recovered through retransmission.

### C. Triggering ACK+Rate Packet

The DRF technique is developed to eliminate the unnecessary transmissions of ACK+Rate packet in order to reduce the additional overhead and energy consumption bared by the intermediate nodes. In this paper, through detailed arguments it is told that the triggering of ACK+Rate packet in response to expiry of epoch timer causes serious performance problems. If the epoch time period is set to 1 second, then the rate changes that occur with in 1 second could not be informed to the sender. This causes harmful effects such as reduction in throughput and under utilization of network resources. To overcome this drawback, DRF sends ACK+Rate packet when ever there is a rate changes, rather than sending ACK+Rate packet for each and every expiry of epoch timer. If ACK+Rate packet is transmitted for every rate change, then more traffic will be generated in the network which causes high energy consumption in the intermediate node and reduction in the throughput due to the contention of data and ACK+Rate packet in the forward and reverse path respectively. If ACK+Rate packet is transmitted after a ±100% rate change(i.e., if the current rate is 4, then the next ACK+Rate packet will be transmitted only if rate becomes 8) than the previous rate, then the sender will choose an inappropriate rate which does not suits exactly with the characteristics of the network at that particular instant. Therefore, in order to find out the optimal time to transmit ACK+Rate packet an experiment is conducted in ns2 simulator. The simulation set up is discussed in section 5.1. The source node and destination node are randomly chosen. Throughput is analyzed in various angles. Transmission of ACK+Rate packets with respect to ± 15%, ±25%, ±35%, ±50%, ±65%, ±75% rate changes than the previous rate is analyzed. Throughput in pedestrian mobile topology (1m/s), throughput in slow speed mobile topology (20m/s), and throughput in high speed mobile topology (30m /s) is found out for 1flow, 5flow and 25flow. The results are shown in Table1. The results are rounded to nearest integer. The average throughput for pedestrian, slow speed and high speed mobile topology for ±15%, ±25%, ±35%, ±50%, ±65%,±75% rate changes in a network load of 1 flow are 539, 529, 468, 366, 278, 187 respectively. From the result it is clear that the throughput for ±15%, ±25% and ±35% are greater than that of ±50%, ±65%, ±75%. Therefore, it can be concluded that ACK+Rate packet may be triggered for transmission if there is ±15%, ±25% or ±25% of rate change than the previous rate change.

TABLE 1:    THROUGHPUT STATISTICS OF DIFFERENT MOBILITY FOR VARIOUS LOADS.

| Transmission of ACK+Rate packet with respect to % of rate change | Throughput in pedestrian topology | | | Throughput in slow speed topology | | | Throughput in high speed topology | | |
|---|---|---|---|---|---|---|---|---|---|
| | 1 flow | 5 flow | 25 flow | 1 flow | 5 flow | 25 flow | 1 flow | 5 flow | 25 flow |
| ±15 | 635 | 214 | 46 | 551 | 157 | 43 | 432 | 141 | 38 |
| ±25 | 629 | 198 | 39 | 537 | 146 | 39 | 421 | 133 | 32 |
| ±35 | 522 | 153 | 31 | 492 | 114 | 31 | 392 | 111 | 24 |
| ±50 | 424 | 132 | 27 | 371 | 91 | 24 | 304 | 92 | 17 |
| ±65 | 367 | 123 | 21 | 254 | 68 | 19 | 213 | 63 | 14 |
| ±75 | 217 | 105 | 19 | 184 | 54 | 15 | 161 | 43 | 8 |

TABLE 2:    STATISTICS OF ACK+RATE PACKET TRANSMISSION

| Transmission of ACK+Rate packet with respect to % of rate change | Number of ACK+Rate packet transmitted for simulation time of 100 second and a network load of 1 flow. | | |
|---|---|---|---|
| | Pedestrian Topology | Slow Speed Topology | High Speed Topology |
| ±15 | 43 | 56 | 69 |
| ±25 | 64 | 73 | 84 |
| ±35 | 76 | 85 | 94 |

Transmission of ACK+Rate packet from the rate change of ±15%, ±25% and ±25% than the previous rate is analysed as shown in table 2. The appropriate percent of rate change so as when to trigger the ACK+Rate packet must be chosen. From the results shown in Table 2, it   clear that the number of ACK+Rate packet in ±25% rate change is lower than ±35% rate change and slightly higher than ±15% rate change. Therefore in DRF the method of triggering ACK+Rate packet, whenever there is ±25% rate change than the previous rate is adopted.

## V.    PERFORMANCE EVALUATION

This section presents the evaluation of DRF and ATP considering the aspects such as energy consumption, change in rate with respect to time for various mobility. The performance of DRF is compared with ATP. The reason for the comparison with ATP is that it is a well known and widely accepted rate based transport protocol. Since the scope of this paper limits to rate based transport protocol, other version of TCP is not considered for comparison

### A.    Simulation Setup

Simulation study was conducted using ns2 network simulator. A mobile topology of 500m X 500m grid consisting of 50 nodes is used. Radio transmission range of each node is kept as 200 meters. The interference range is kept as 500 meters. Channel capacity is chosen as 2 Mbit/sec. channel delay is fixed as 2.5μs. Dynamic source routing and IEEE 802.11b is used as the routing and MAC protocol. FTP is used as the application layer protocol. The source and destination pairs are randomly chosen. The effect of load on the network is studied with 1,5 and 25 flows. The performance of DRF is evaluated and compared with ATP.

### B.    Instantaneous Rate Dynamics

Instantaneous rate dynamics refers to change in rate (packets/sec) with respect to time. Fig.1 shows the result of change in rate for various mobility 1m/s, 10m/s, 30m/s, 50m/s. When mobility is 1 m/s, rate change occurs 3 to 4 times. The average rate of rate changes is 210. When mobility is 50 m/s, frequent rate change occurs. The average rate of rate changes is 225.4. While the mobility is 50 m/s, the maximum deviation of rate change with respect to average value ranges from 50 to 60. From this observation it can be concluded that mobility is directly proportional to the change in rate i.e., whenever mobility increases the rate change also increases.

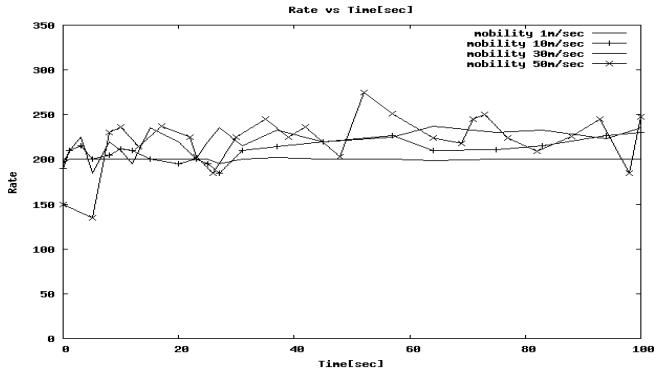

Figure 1. Rate Vs Time

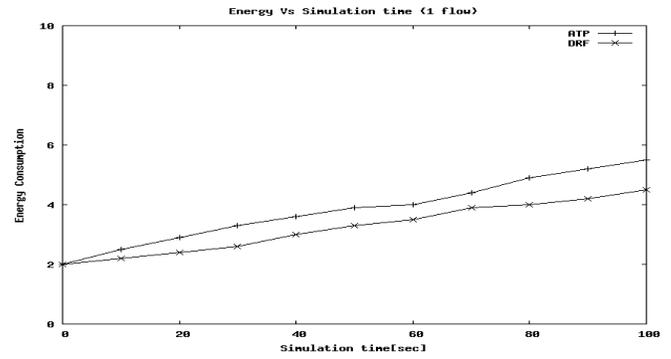

(a)

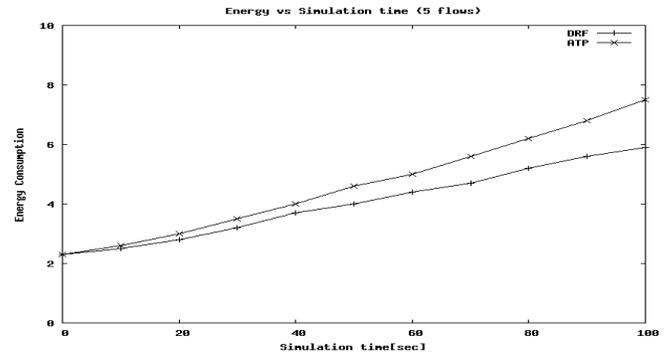

(b)

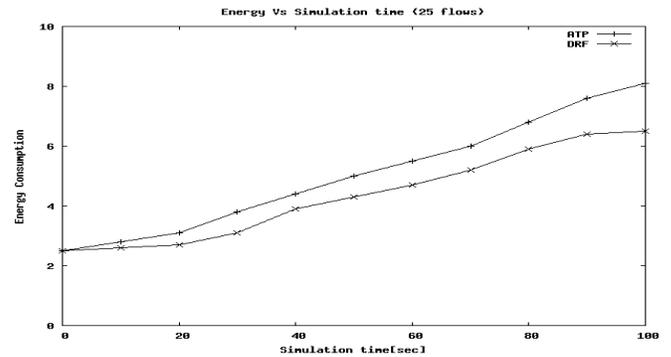

(c)

Figure 2. Energy Consumption in Pedestrian Mobile Topology (a) [1flow] (b) [5flow], (c) [25flow].

### D. Energy Consumption

In this section, the energy consumption in intermediate node is examined for ATP and DRF. The energy model implemented in the ns2 simulator [22] is used in simulation. A node starts with initial energy level that is reduced whenever the node transmits, receives a packet. The total amount of energy E(n) consumed at a given node n is represented as

$E(n) = E_t(n) + E_r(n)$

Where $E_t$ and $E_r$ denote the amount of energy spended by transmission, reception respectively. Energy/ Bit ratio i.e., the energy consumption per bit is computed as follows

$e = n*p*8/es$

Where n is the number of packet transmitted, p is the size of the packet in bytes and es is the energy spend by the node in joules.

### E. Performance in pedestrian (1 m/s) mobile topology

In pedestrian mobile topology, the rate change with respect to time is low as compared to the slow speed and high speed mobile topology. Snapshot of energy consumption due to the transmission of ACK+Rate packet in DRF and ATP are presented for single flow, 5 flows, and 25 flows in Fig. 2 a, b, c respectively. Here the focus is only to high light the energy consumption of ATP and DRF. Hence other variants of TCP and other rate based transport protocols are not considered. The average energy consumption of ATP are 3.84, 4.65, 5.05 for 1 flow, 5 flow, 25 flow respectively. The average energy consumption of DRF are 3.24, 4.01, 4.35 for 1 flow, 5 flow, 25 flow respectively. In ATP since the ACK+Rate packet is transmitted for every epoch period (say 1 second), the number ACK+Rate packet transmitted is high. Therefore the energy consumption in ATP is higher as compared to DRF. Energy consumption in DRF is low because ACK+Rate is transmitted to the sender only when there is rate change. In pedestrian topology since the mobility is low ( 1m/s), the rate change is also low so the number of ACK+ Rate packet transmission is minimum. Therefore the energy consumption for both ATP and DRF is minimum as compared to slow speed and high speed mobile topology.

### F. Performance in slow speed mobile topology

Snapshot of energy consumption due to the transmission of ACK + Rate packet in DRF and ATP are presented for single flow, 5 flows, and 25 flows in Fig.3 a, b, c respectively. The average energy consumption of ATP are 3.92, 4.75, 5.12 for 1 flow, 5 flows, 25 flows respectively. The average energy consumption of DRF are 3.75, 3.94, 4.24 for 1 flow, 25 flow respectively. In slow speed mobile topology the speed of mobility is 10m/s. The average energy consumption in slow speed mobile topology is greater than that of the results obtained in pedestrian mobile topology. According to the results shown in Fig. 1. as the speed increases the change in rate with respect to time also increases. Therefore energy consumption in both ATP and DRF is higher in slow speed topology than that of pedestrian mobile topology. But comparing ATP and DRF, ATP has high energy consumption than DRF, this is due to the transmission of ACK+Rate packet based on epoch timer expiry. This is the indication to use DRF technique in a topology where mobility speed varies from 1 m/s to 10 m/s. DRF causes reduction of energy consumption and there by increases the network life time which is critical issue.

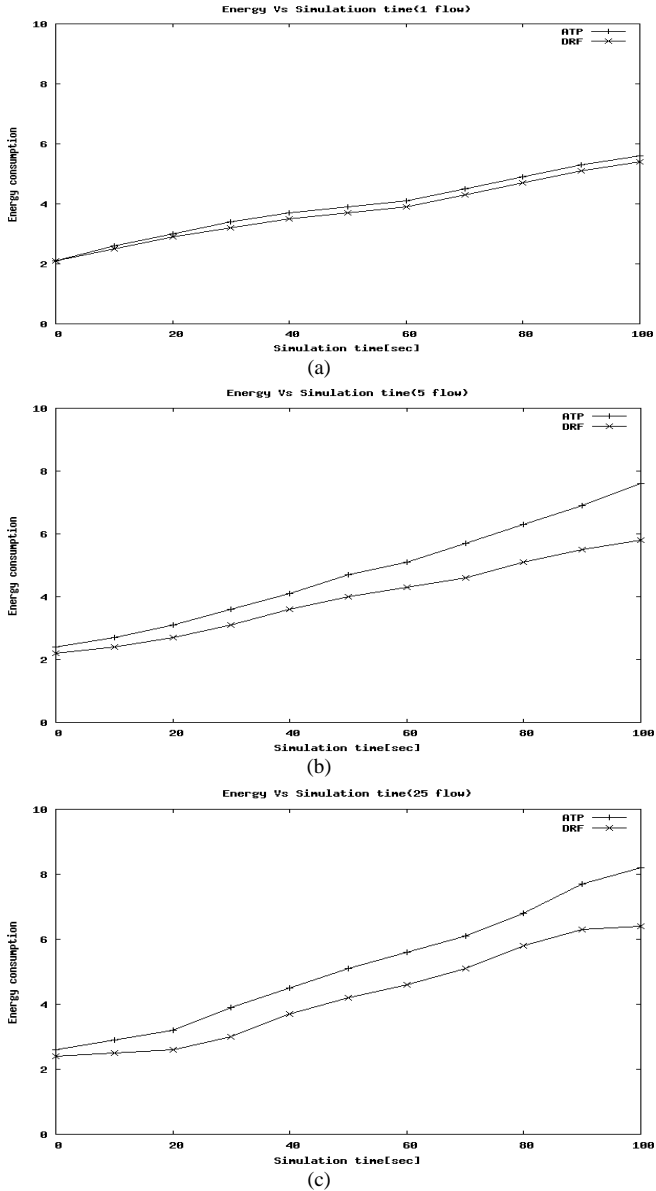

Figure 3. Energy Consumption in slow speed mobile topology (a) [1flow], (b) [5flow],(c) [25flow].

### G. Performance in high speed mobile topology

Snapshot of energy consumption for high speed mobile topology of ATP and DRF are presented for single flow, 5 flows and 25 flows in Fig.4 a, b, c respectively. The average energy consumption of ATP are 4.15, 4.85, 5.33 for 1 flow, 5 flows and 25 flows respectively. The average energy consumption of DRF are 4.89, 5.6, 6.23 for 1 flow, 5 flow, 25 flow respectively It is seen that the energy consumption of DRF is greater than that of ATP. This is because as mobility increases rate change also increases. In DRF the ACK+Rate packet is transmitted whenever there is a rate change. Since rate changes are higher, number of ACK+ Rate packets transmitted is higher. The result shows that within 1 second 5 to 6 ACK+Rate packets are transmitted. This raises the energy consumption. In case of ATP ACK+Rate packet is transmitted for each and every epoch period (say 1 second) in contrast, DRF send ACK+Rate packet whenever there is a rate change.

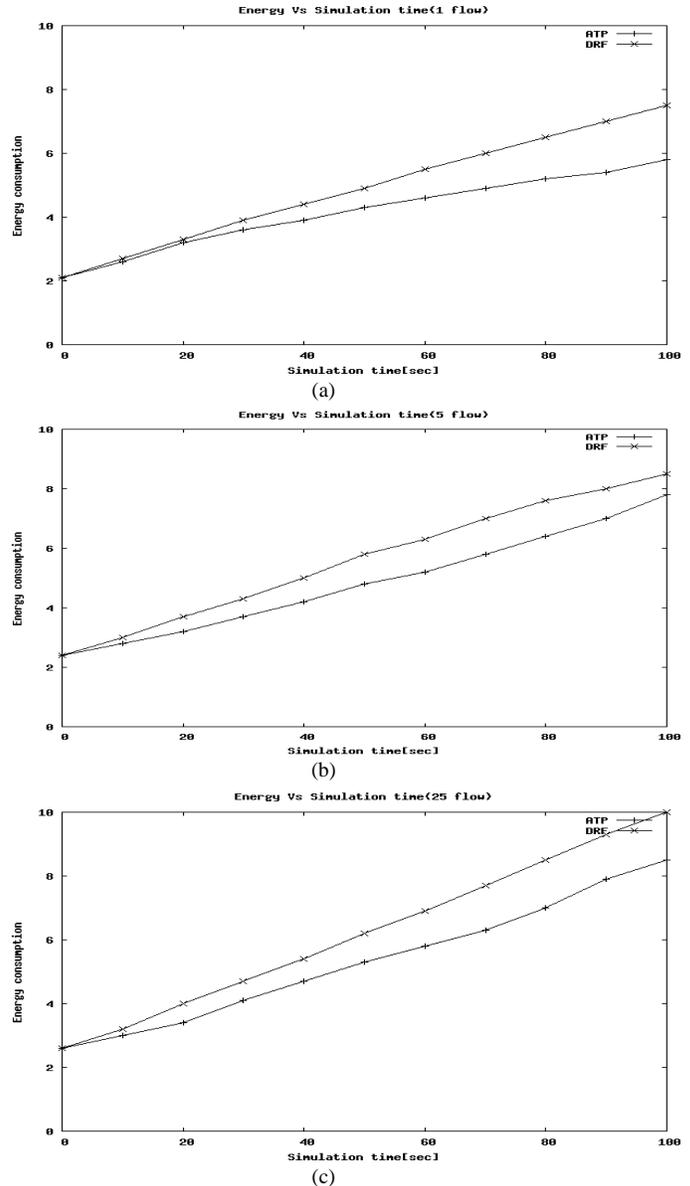

Figure 4. Energy Consumption in High speed mobile topology (a) [1flow], (b) [5flow], (c) [25flow].

The results clearly reveals that the DRF technique reduces energy consumption and increases the life time of a node in very slow speed and slow speed mobility. But in high speed mobility topology DRF consumes slightly higher energy than ATP. Therefore it is clear that DRF technique can be deployed in an ad hoc network where the mobility speed does not exceed 20 m/s.

## VI.  CONCLUSION

The Dynamic Rate Feedback (DRF) strategy aims to minimize the contention between data and ACK+Rate packets by transmitting as few ACK+Rate packet as possible. This mechanism is self adaptive and it is feasible and optimal to use in ad hoc networks whose mobility is restricted to 20m/s. Through simulation it is found that the frequency of change in rate is directly proportional to the mobility. In DRF technique, it is adopted that the ACK+Rate is transmitted only, whenever there is change in rate of ±25% than the previous rate. The simulation result showed that the DRF can outperform ATP, the well known rate based transport protocol for ad hoc network in terms of energy consumption in intermediate node. This technique is easy to deploy since the changes are limited to the end node (receiver) only. It is important to emphasize that in DRF the semantics of ATP is retained but the performance is improved.

## AUTHORS PROFILE

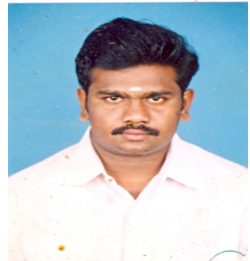

*P.Ganesh kumar* received the bachelor's degree in Electrical and Electronics Engineering from Madurai Kamaraj University in 2001, and Master's degree in Computer science and Engineering with distinction from Bharathiar University in 2002. He is currently working towards the  PhD degree in Anna university Chennai. He has 7 years of teaching experience in Information Technology. He is a member of IEEE, CSI and ISTE. He had published several papers in International journal, IEEE international conferences and national conferences. He authored a book "Component based Technology". His area of interest includes Distributes systems, Computer Network and Ad hoc network.

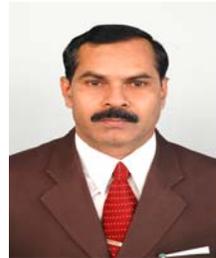

*Dr. K.Thyagarajah* received the bachelor's degree in Electrical Engineering and master's degree in Power systems from Madras university. He received doctoral degree in  Power Electronics and AC Motor Drives from Indian Institute of science, Banglore in 1993. He has  30 years of experience in teaching and research. He is a senior member of IEEE. He is a senior member of various bodies such as ISTE, Institution of engineers, India etc. He is syndicate member in Anna university Chennai and in Anna University Tiruchirapalli. He is member of board of studies in various universities in India. He has published more than 50 papers in various national and international referred journals and conferences. He authored a book "Advanced Microprocessor". His area of interest includes Network Analysis, Power electronics, Mobile computing, Ad hoc networks.